\documentclass[prl,twocolumn,amssymb]{revtex4}
\usepackage{graphicx}% Include figure files
\usepackage{dcolumn}% Align table columns on decimal point
\usepackage{bm}% bold math
\usepackage{wasysym}
\newcommand{\beq}{\begin{equation}}
\newcommand{\eeq}{\end{equation}}
\newcommand{\beqn}{\begin{eqnarray}}
\newcommand{\eeqn}{\end{eqnarray}}

\begin{document}
\title{Low Energy Effective Field Theories of Sp(4) Spin systems}
\author{Cenke Xu}
\affiliation{Department of Physics, Harvard University, Cambridge,
MA 02138}
%\author{Yang Qi}
%\affiliation{Department of Physics, Harvard University, Cambridge,
%MA 02138}
\date{\today}

\begin{abstract}

We study the classical and quantum phase transitions of Sp(4) spin
systems on three dimensional stacked square and triangular
lattices. We present general Ginzburg-Landau field theories for
various types of Sp(4) spin orders with different ground state
manifolds such as CP(3), $S^7/Z_2$, Grassmann manifold $G_{2,5}$,
$G_{2,6}$ and so on, based on which the nature of the classical
phase transitions are studied, and a global phase diagram is
presented. The classical phase transitions close to quantum phase
transitions toward spin liquid states are also discussed based on
renormalization group (RG) flow. Our results can be directly
applied to the simplest Sp(4) and SU(4) Heisenberg models which
can be realized using spin-3/2 atoms and Alkaline earth atoms
trapped in optical lattice.

%On the stacked square lattice, by tuning one parameter the system
%undergoes two transitions separating a Neel phase, a photon phase,
%and a plaquette ordered phase. At finite temperature we propose
%two separate transitions in the Neel order side, with one O(3)
%transition followed by an O(5) transition at higher temperature.
%One the stacked triangular lattice, a transition to a $Z_2$
%quantum spin liquid is realized by tuning one parameter in the
%nearest neighbor Heisenberg model. At finite temperature there are
%two possible scenarios: 1st, one 3d O(8) transition followed by a
%3d Ising transition at higher temperature with an intermediate
%classical $Z_2$ spin liquid; 2nd, one ``coupled O(3)" transition
%and an O(5) transition at higher temperature with an intermediate
%nematic phase. In the vicinity of the quantum critical point the
%first scenario is valid.

\end{abstract}
 \maketitle

\section{1, introduction}

For decades condensed matter physicists have been actively
studying the spin systems with large symmetries such as SU(N) and
Sp(N)
\cite{arovas1988,sachdev1989,sachdev1990,sachdev1991,affleck1987,affleck1991,
brad1988,brad1989,coleman2009,coleman2008}, mainly motivated by
the fact that under large-N generalization the semiclassical spin
order with spin symmetry breaking is weakened, and even vanishes
completely beyond certain critical $\mathrm{N}_c$. A good example
is the SU(N) Heisenberg model on square lattice with fundamental
and conjugate fundamental representation on two sublattices (FCF
Heisenberg model), which for $\mathrm{N} > 4$ is quantum
disordered, and for $\mathrm{N} \leq 4$ the ground state
spontaneously breaks the SU(N) symmetry, with ground state
manifold (GSM) $\mathrm{CP(N-1)}$
\cite{sachdev1990,gmzhang2001,kawashima2003}. Very recently it was
proposed that, without fine-tuning any parameter, the SU(N) spin
systems with N as large as 10 can be realized by alkaline earth
atoms trapped in optical lattice \cite{alkaline}, so the large-N
spin system is no longer merely theoretical toy. Many previous
works showed that for the special value $\mathrm{N} = 4$, the
Sp(4) symmetry can be realized with spin-3/2 fermionic atoms, and
when the spin-0 and spin-2 $s-$wave scattering lengths are equal,
the system has an even larger SU(4) symmetry \cite{wu2003}.
Motivated by these observations, quantum magnetism based on the
spin-3/2 atoms has been actively studied
\cite{wu2003,wu2005,wu2005a,wu2006b,xuwu2008,tu2006,tu2007}.

%and it was shown by Schwinger boson formalism since early 1990s
%that for $1/N
%> 0.19$ the ground state of this model is Neel order with ground
%state manifold (GSM) $\mathrm{CP(N-1)}$, otherwise the ground
%state is classically disordered, with long range dimerized
%crystalline order induced by quantum fluctuation. Recent numerical
%results confirmed this result, and showed that for $N > 4$, the
%ground state is disordered, and for $N = 4$ there is a tiny Neel
%order moment, which is consistent with previous Schwinger boson
%calculations.

Although the GSM of the ordered state of SU(N) FCF Heisenberg
model on square lattice has been identified as $\mathrm{CP(N-1)}$
long ago, a detailed Ginzburg-Landau (GL) field theory for this
ordered state has not been thoroughly studied. A GL theory of this
state can answer the following question: Suppose the SU(N)
Heisenberg model is defined on the 3d cubic lattice with
$\mathrm{CP(N-1)}$ GSM, what is the finite temperature transition
between this ordered phase at low temperature and a disordered
phase at high temperature? For $N = 2$, this question is fairly
simple, because $\mathrm{CP(1)} = S^2$, the finite temperature
transition is no more than one single 3d O(3) transition. For
larger-$N$ cases, the question is complicated by the fact that
$\mathrm{CP(N-1)}$ manifold does not have a general simple
parametrization as $\mathrm{CP(1)}$. The standard way to
parameterize the $\mathrm{CP(N - 1)}$ manifold is to treat it as
$N$ component of complex boson coupled with U(1) gauge field,
while keeping the SU(N) global symmetry of the action, but this
parametrization of $\mathrm{CP(N - 1)}$ manifold fails to describe
the finite temperature phase transition, which will be discussed
in the next section. Therefore we need to write down a GL theory
based only on the physical observable order parameters.

In the current work we will focus on the case with $N  = 4$ and
discuss the finite temperature phase transition of system with
CP(3) GSM. One sample system which has CP(3) GSM is the Sp(4)
Heisenberg model on bipartite lattice with one particle per site:
\beqn H = \sum_{<i,j>} J_1 \Gamma^{ab}_i\Gamma^{ab}_j -
J_2\Gamma^a_i\Gamma^a_j. \label{model}\eeqn $\Gamma^a$ with $a =
1...5$ are five $4 \times 4$ Gamma matrices, and $\Gamma^{ab} =
\frac{1}{2i}[\Gamma^a, \Gamma^b]$ are 10 generators of Sp(4)
$\sim$ SO(5) group. Here we choose the following standard
convention of Gamma matrices: \beqn \Gamma^a = \mu^z \otimes
\sigma^a, \ a = 1, 2, 3, \ \ \Gamma^4 = \mu^x \otimes \mathbf{1} ,
\ \Gamma^5 = \mu^y \otimes \mathbf{1}. \eeqn For arbitrary $J_1$
and $J_2$ this system has Sp(4) symmetry, while when $J_1 = J_2$
this model is equivalent to the SU(4) FCF Heisenberg model
\cite{wu2006b}. $J_1$ and $J_2$ can be tuned with spin-0 and
spin-2 $s-$wave scattering lengths of spin-3/2 cold atoms
\cite{wu2006b}. Our formalism suggests that for a Sp(4) spin
system on the 3d cubic lattice with GSM CP(3), depending on the
ratio $J_1/J_2$ the classical phase diagram can have different
scenarios. The most interesting scenario is the region $J_2 > J_1$
in model Eq. \ref{model}, at finite temperature there are two
transitions, with one 3d O(5) transition followed by a 3d O(3)
transition at lower temperature. On stacked triangular lattice, it
was shown that the Sp(4) Heisenberg model Eq. \ref{model} has
$\sqrt{3}\times \sqrt{3}$ spin order with GSM $S^7/Z_2$
\cite{xuyang2008}. At finite temperature again there can be two
transitions, with one 3d O(5) transition followed by a ``coupled"
O(3) transition. Besides CP(3) and $S^7/Z_2$, many other spin
symmetry breaking semiclassical states of Sp(4) spins with
different GSM can exist, especially for half-filled (2-particle
per site) system, which will also be discussed in this work.

This paper is organized as follows: In section II, we will study
the GL theory of the Neel and $\sqrt{3}\times \sqrt{3}$ phases of
the Sp(4) spin system on three dimensional lattices, and a global
phase diagram is presented. Sp(4) spin states with other GSM such
as Grassmann manifold $G_{2,5}$, $G_{2,6}$ and
$\mathrm{SO(5)/SO(3)}$ will also be discussed, with applications
to half-filled Sp(4) spin models. Our GL theory can also be used
to distinguish different GSMs with the same dimension and similar
quotient space representation. In section III we will study the
classical phase transitions close to quantum phase transitions
between ordered and spin liquid phases. In section IV we will
briefly discuss a more exotic manifold, the ``squashed $S^7$" and
its potential to be realized in Sp(4) spin systems.

%Our formalism also shows that the GSM $\mathrm{SO(5)/[SO(3)\times
%SO(2)]}$ of half-filled spin-3/2 cold atom in optical lattice is
%actually different from CP(3), although it is also 6 dimensional.

%Ref.  also showed that the GSM of the $\sqrt{3}\times \sqrt{3}$
%phase of the Sp(4) Heisenberg model on the triangular lattice has
%GSM $S^7/Z_2$. In our current work we will also study the finite
%temperature transition of the $\sqrt{3}\times \sqrt{3}$ phase of
%the Sp(4) Heisenberg model on the 3d stacked triangular lattice. A
%GL theory of this state still suggests two transitions, with an
%O(5) transition at higher temperature and a ``coupled" O(3)
%transition at lower temperature.

\section{2, GL theories for Sp(4) spin systems}

%at finite temperature the transition between low temperature
%ordered phase with CP(3) GSM and high temperature disordered phase
%is suffering from temperature generated monopoles, which is
%relevant at the critical point described by CP(3) field theory
%(\ref{cpn})

\subsection{A, Collinear Phases}

Let us now consider the Sp(4) Heisenberg model Eq. \ref{model} on
the 3d cubic lattice with 1 particle per site. On the 2d square
lattice, both analytical and numerical results conclude that at
the special point $J_1 = J_2$ with enlarged SU(4) symmetry, the
ground state of this model has semiclassical order
\cite{sachdev1990,gmzhang2001,kawashima2003}, with GSM CP(3),
which extends into a finite range of the phase diagram tuned by
$J_2/J_1$ \cite{xuyang2008}. The semiclassical order is expected
to be stable with the third direction unfrustrated interlayer
coupling. In this Neel phase, both $\Gamma^{ab}$ and $\Gamma^a$
are ordered. For instance, we can take the trial single site state
$|\psi \rangle = (1, 0, 0, 0)^t$, and it is trivial to see that it
has nonzero $\Gamma^{3}$, $\Gamma^{45}$ and $\Gamma^{12}$.

As already mentioned in the introduction, the standard way to
parameterize the $\mathrm{CP(N - 1)}$ manifold is to treat it as N
component of complex boson coupled with U(1) gauge field, while
keeping the global spin symmetry of the action: \beqn L =
|(\partial_\mu - iA_\mu)z|^2 + r|z|^2 + g (|z|^2)^2 + \cdots
\label{cpn}\eeqn This action is written down based on the fact
that \beqn \mathrm{CP(N - 1)} = S^{2N - 1}/\mathrm{U(1)}. \eeqn
Here $S^{2N-1}$ represents the GSM of the condensate of N
component of complex boson, and U(1) represents the U(1) gauge
field $A_{\mu}$. However, at finite temperature, a simple
$\mathrm{CP(N - 1)}$ model in Eq. \ref{cpn} on three spatial
dimension would lead to a wrong transition, because this model
describes a transition between the ordered phase and a photon
phase. However, finite temperature induces finite density of
monopoles of $A_\mu$, which will change the photon propagator at
long scale. The disordered phase is generically identical to the
``confined phase" with monopole proliferation and no lattice
symmetry breaking $i.e.$ the monopoles without Berry phase.
Therefore the action (\ref{cpn}) should be supplemented with the
``featureless" monopole, which is relevant at least for small N at
the critical point $r = 0$. For $\mathrm{N} = 2$, the ``trivial"
monopole drives the CP(1) model to the O(3) universality class,
but for larger N there is no such simple relation. Therefore the
$\mathrm{CP(N - 1)}$ model plus monopole does not tell us much
about the nature of the transition in general, and we need another
convenient way to describe the $\mathrm{CP(N - 1)}$ manifold.

Therefore, to describe the GSM and transition we need to introduce
a linear sigma model at $4 - \epsilon$ dimension with gauge
invariant order parameters, in the form of $z^\dagger_\alpha
z_\beta$. There are in total 15 independent bilinears of this
form, which can be simply rewritten as the following 5 component
vector and 10 component adjoint vector: \beqn && \phi^{ab} =
z^\dagger \Gamma_{ab}z, \ \ \phi^{a} = z^\dagger \Gamma_{a}z \cr
\cr && \sum_{a, b} \phi^{ab}\phi^{ab} \sim \sum_{a}
\phi^{a}\phi^{a} \sim (|z|^2)^2. \eeqn The complex bosonic field
$z_\alpha$ are the low energy Schwinger bosons of Sp(4) spin
system. In Ref. \cite{xuyang2008}, it was shown that in the Neel
order $\phi^{ab}$ is the staggered order $(-1)^i\Gamma^{ab}$ while
the O(5) vector $\phi^a$ is the uniform order $\Gamma^a$, which
can be naturally expected from Eq. \ref{model}, when $J_1$ and
$J_2$ are both positive. However, $\phi^{ab}$ and $\phi^a$ are not
independent vectors, because the Sp(4) symmetry of the system
allows for coupling between these two vectors in the free energy,
which can be manifested by the following identities: \beqn \sum_{a
= 1}^5\phi^a\phi^a &=& 2(|z|^2)^2, \ \ \ \sum_{a} \phi^a N^a =
2(|z|^2)^3, \cr \cr N^a &=& \epsilon_{abcde}\phi^{bc}\phi^{de}.
\eeqn $\epsilon_{abcde}$ is the five dimensional antisymmetric
tensor. Also, the five $\Gamma^a$ matrices are all constructed by
bilinears of the spin-3/2 operators, while $\Gamma^{ab}$ are
constructed by linear and cubics of the spin operators
\cite{wu2003,wu2006b}. Therefore $\vec{\phi}$ is time reversal
even, and identical to the nematic O(5) vector $N^a =
\epsilon_{abcde}\phi^{bc}\phi^{de}$ in the ordered state of the
$\mathrm{CP(3)}$ model with $|z|^2 = 1$.

Now we can write down a classical GL theory for Sp(4) spin system
with CP(3) GSM: \beqn F &=& \sum_{ab,\mu}(\nabla_\mu \phi^{ab})^2
+ (\nabla_\mu \phi^a)^2 + r_1 (\phi^{ab})^2 + r_2 (\phi^a)^2
\cr\cr &+& \gamma\epsilon_{abcde}\phi^a\phi^{bc}\phi^{de} + g\{
\sum_{ab}(\phi^{ab})^2 + \sum_a(\phi^a)^2\}^2 \cr\cr &+& \cdots
\label{GL}\eeqn The ellipses include all the other terms allowed
by Sp(4) global symmetry. When $r_1 = r_2$, this free energy is
SO(6)$\sim$SU(4) invariant, which corresponds to the point $J_1 =
J_2$, where the model is equivalent to the SU(4) FCF Heisenberg
model on the cubic lattice. We can also view the adjoint vector
$\phi^{ab}$ as an O(10) vector which originally should form a GSM
$S^9$, and the cubic term $\gamma$ makes the ten component vector
$\phi^{ab}$ align in a 6 dimensional submanifold of $S^9$ where
the O(5) vector $\phi^a \sim \epsilon_{abcde}\phi^{bc}\phi^{de}$
is maximized.

\begin{figure}
\includegraphics[width=2.5in]{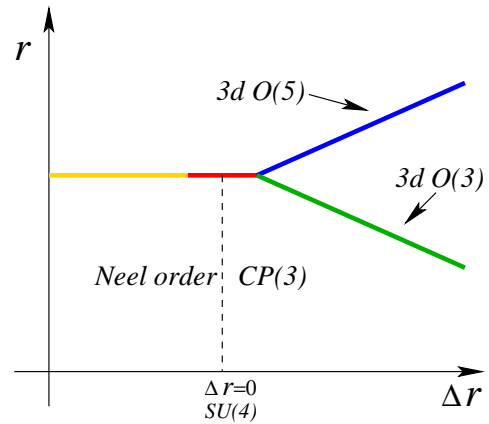}
\caption{The phase diagram of GL theory Eq. \ref{GL}, plotted
against $r = r_1 + r_2$ and $\Delta r = r_1 - r_2$. The red line
is a first order transition, the blue line is a 3d O(5)
transition, the green line is a 3d O(3) transition. The golden
line is a second order transition at the mean field level, the
true nature of the transition can be obtained by a detailed RG
calculation for Eq. \ref{GL3} with $\gamma_2 < 0$. A similar phase
diagram can be applied to Eq. \ref{GL4} for the stacked triangular
lattice, with the green line representing a coupled O(3)
transition described by Eq. \ref{n1n2}.} \label{sp4phasedia}
\end{figure}

A global mean field phase diagram can be plotted against $r = r_1
+ r_2$ and $\Delta r = r_1 - r_2$, as shown in Fig.
\ref{sp4phasedia}. The parameter $r$ is tuned by temperature, and
$\Delta r$ is tuned by $\Delta J = J_1 - J_2$, which is evident
with the observation that $\Delta r = 0$ corresponds to the same
SU(4) point $\Delta J = 0$ and both finite $\Delta r$ and $\Delta
J$ violate the SU(4) symmetry. There are three different regions
in the phase diagram. Close to the SU(4) point $\Delta r = 0$, the
cubic term $\gamma$ drives a first order transition at the mean
field level, with both $\langle\phi^{ab}\rangle$ and
$\langle\phi^a\rangle$ jump discontinuously. The first order
transition extends to a finite region in the phase diagram. The
second region of the phase diagram has $\Delta r < 0$ ($ J_1 >
J_2$), here $\phi^{ab}$ wants to order before $\phi^a$, but due to
the $\gamma$ term in the free energy (\ref{GL}), the order of
$\phi^{ab}$ implies order of $\phi^a$. Therefore in this region
the phase transition can be safely described by a free energy in
terms of only $\phi^{ab}$, after integrating out $\phi^a$: \beqn
F_2 &=& \sum_{ab,\mu}(\nabla_\mu \phi^{ab})^2 + r(\phi^{ab})^2 +
\gamma_2 \sum_a (\epsilon_{abcde}\phi^{bc}\phi^{ed})^2 \cr\cr  &+&
g_2 (\sum_{ab}(\phi^{ab})^2)^2 + \cdots \label{GL3}\eeqn Here
$\gamma_2 < 0 $ to make sure the ground state wants to maximize
$\phi^a$. We can treat $\gamma_2$ as a perturbation at the 3d
O(10) transition, and a coupled renormalization group flow of
$\gamma_2$ and $g_2$ will determine the fate of the transition.

The third region is $\Delta r > 0$ ($ J_1 < J_2$), now $\phi^a$
tends to order before $\phi^{ab}$, and there are in general two
separate second order transitions at finite temperature, with
$\phi^a$ orders first. The transition of $\phi^a$ is a three
dimensional O(5) transition. After the ordering of $\phi^a$, the
symmetry of the system breaks down to O(4). Let us take the
expectation value of $\vec{\phi}$ as $\langle\vec{\phi}\rangle =
(\sigma, 0 , 0, 0, 0 )$, the coupling between $\phi^a$ and
$\phi^{ab}$ in free energy (\ref{GL}) reads: \beqn
\epsilon_{abcde}\phi^a\phi^{bc}\phi^{de} = \sigma
(\phi^{23}\phi^{45} - \phi^{24}\phi^{35} + \phi^{25}\phi^{34}).
\label{quadra} \eeqn Now one can diagonalize the quadratic part of
the Eq. \ref{GL} and Eq. \ref{quadra}, the eigenmodes are
characterized by the representation of the residual $\mathrm{O(4)}
\simeq \mathrm{SU(2) \times SU(2)}$ symmetry. The residual O(4)
symmetry group is generated by 6 matrices $\Gamma^{ab}$ with $a, b
\neq 1$. The two SU(2) normal subgroups of O(4) are generated by
matrices $- \Gamma^{23} + \Gamma^{45}$, $\Gamma^{24} +
\Gamma^{35}$, $- \Gamma^{25} + \Gamma^{34}$ (denoted as subalgebra
$\mathrm{su(2)}_A$) and $ \Gamma^{23} + \Gamma^{45}$, $-
\Gamma^{24} + \Gamma^{35}$, $\Gamma^{25} + \Gamma^{34}$ (denoted
as subalgebra $\mathrm{su(2)}_B$) respectively. We will decompose
the 10 component vector $\phi^{ab}$ based on the representation of
the $\mathrm{su(2)}_A$ and $\mathrm{su(2)}_B$ algebras, different
representations will have different eigenvalues: \beqn \vec{Q}^i
(i = 1 \cdots 4) &=& (\phi^{12}, \ \phi^{13}, \ \phi^{14},
\phi^{15}), \cr \cr \mathrm{eigenvalue} &:& r , \ \ \
\mathrm{Representation}: \mathrm{O(4)} \ \mathrm{vector}; \cr\cr
\vec{T}^i_A (i = 1, 2, 3) &=& ( - \phi^{23} + \phi^{45}, \
\phi^{24} + \phi^{35}, \ - \phi^{25} + \phi^{34}), \cr \cr
\mathrm{eigenvalue} &:& r - \gamma\sigma, \ \ \
\mathrm{Representation}: (1, 0); \cr\cr \vec{T}^i_B (i = 1, 2, 3)
&=& ( \phi^{23} + \phi^{45}, \ - \phi^{24} + \phi^{35}, \
\phi^{25} + \phi^{34}), \cr \cr \mathrm{eigenvalue} &:& r +
\gamma\sigma, \ \ \ \mathrm{Representation}: (0, 1). \eeqn Here
$\vec{T}_A$ and $\vec{T}_B$ transform as vectors of
$\mathrm{SU(2)}_A$ and $\mathrm{SU(2)}_B$ respectively. Notice
that although $\mathrm{SU(2)}_A$ and $\mathrm{SU(2)}_B$ are both
normal subgroups of the SO(4) after the order of $\phi^a$, neither
of them can be normal subgroup of the original SO(5) group,
because SO(5) group is a simple group while SO(4) is a semisimple
group.

If $\gamma\sigma > 0$, $\vec{T}_A$ has the lowest eigenvalue, so
the O(3) vector $\vec{T}_A$ will order after $\phi^a$. The main
question is which universality this transition belongs to. Since
$\vec{Q}$ and $\vec{T}_B$ are massive and only have short range
correlation at the transition of $\vec{T}_A$, integrating out them
will not induce any critical behavior for $\vec{T}_A$, and hence
the Goldstone mode of $\phi^a$ after its ordering is the biggest
concern. The Goldstone mode $(0, \pi_1, \pi_2, \pi_3, \pi_4)$
forms an O(4) vector, and the Goldstone theorem guarantees its
gaplessness. The simplest coupling one can write down with these
constraints is: \beqn F^\prime \sim (\vec{T}_A)^2 (\nabla_\mu
\vec{\pi})^2. \label{goldstone}\eeqn This term only generates
irrelevant perturbations at the O(3) transition of $\vec{T}_A$
after integrating out $\vec{\pi}$. Notice that couplings like
$(\vec{T}_A)^2 (\vec{\pi})^2$ though preserves the global O(4)
symmetry, violates the Goldstone theorem after integrating out
$\vec{T}_A$, as a mass gap $ \sim \langle \vec{T}^2_A\rangle$ is
induced for $\vec{\pi}$. Therefore now we can safely conclude that
the phase transition of $\vec{T}_A$ is a 3d O(3) transition.
Notice that vector $\vec{T}_B$ and $\vec{Q}$ no longer have to
order at lower temperature, because of the repulsion from ordered
$\vec{T}_A$, due to the quartic terms in Eq. \ref{GL}.

After the ordering of $\vec{T}_A$, the symmetry of the system is
broken down to $\mathrm{SO(2)\times SO(3)}$. The first SO(2)
corresponds to the residual symmetry of $\mathrm{SU(2)}_A$ after
the order of $\vec{T}_A$, and the second SO(3) corresponds to the
$\mathrm{SU(2)}_B$ associated with $\vec{T}_B$, therefore CP(3)
manifold can also be written as quotient space
$\mathrm{SO(5)/[SO(2)\times SO(3)]}$. However, we should be
careful about this formula, because there are two different types
of so(3) or su(2) subalgebras of so(5). Besides the subalgebras
$\mathrm{su(2)}_A$ and $\mathrm{su(2)}_B$ we used earlier, there
is another SU(2) subgroup which is the diagonal subgroup of
$\mathrm{SU(2)}_A \times \mathrm{SU(2)}_B$, we denote this
subgroup as $\mathrm{SU(2)}_V$, which is no longer a normal
subgroup of O(4). The elements in algebra $\mathrm{su(2)}_V$ are
the linear combination of the corresponding elements in
$\mathrm{su(2)}_A$ and $\mathrm{su(2)}_B$: $J^V_i = J^A_i +
J^B_i$.

\begin{figure}
\includegraphics[width=2.5in]{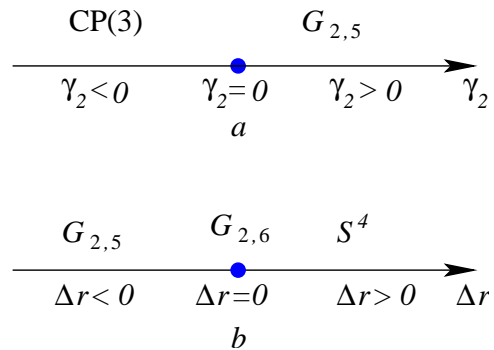}
\caption{The schematic ground state manifold phase diagram of
(Fig. $a$) Eq. \ref{GL3} and (Fig. $b$) Eq. \ref{GLhalffilled},
with $\Delta r = r_1 - r_2$. } \label{halffilled}
\end{figure}

For instance, in the half-filled (2 particles per site) spin-3/2
cold atoms, one can naturally obtain an ordered state with
$\langle (-1)^i \Gamma^{ab} \rangle \neq 0$ but with no order of
$\Gamma^a$ \cite{tu2006,tu2007}, which means that for this case
action Eq. \ref{GL3} is still applicable, while the sign of
$\gamma_2$ is positive $i.e.$ it corresponds to a different
anisotropy of the $S^9$ manifold formed by the adjoint vector
$\phi^{ab}$, which minimizes the vector $\phi^a \sim
\epsilon_{abcde}\phi^{bc}\phi^{de}$ (in contrast to CP(3)) (Fig.
\ref{halffilled}). In this case the GSM can still be written as
$\mathrm{SO(5)/[SO(2)\times SO(3)]}$, but here SO(3) is
$\mathrm{SU(2)}_V$. For instance if
$\langle(-1)^i\Gamma^{12}\rangle \neq 0$, the $\mathrm{SU(2)}_V$
is generated by $\Gamma^{34}$, $\Gamma^{35}$ and $\Gamma^{45}$.
This GSM $\mathrm{SO(5)/[SO(2)\times SO(3)]}$ with
$\mathrm{SO(3)\sim SU(2)}_V$ is called Grassmann manifold
$G_{2,5}$, which is mathematically defined as the set of
2-dimensional planes in 5 dimensional vector space
\cite{nakaharabook}.

The mean field phase diagram for the half-filled Sp(4) system
tuned by the spin-0 and spin-2 $s-$wave scattering lengths is
studied in Ref. \cite{wu2003,tu2006,tu2007}. Besides the phase
with $\langle (-1)^i \Gamma^{ab} \rangle \neq 0$ discussed in the
previous paragraph, there is another typical phase with $\langle
(-1)^i \Gamma^{a} \rangle \neq 0$ and GSM $\mathrm{SO(5)/SO(4)} =
S^4$. These two phases are separated from each other by the SU(4)
point with equal spin-0 and spin-2 scattering lengths, where due
to the enlarged symmetry, the two different orders should have
equal energy \cite{wu2003}. Suppose
$\langle(-1)^i\Gamma^{12}\rangle$ is nonzero at this SU(4) point,
now the residual symmetry of this order is generated by
$\Gamma^{12}$, $\Gamma^{34}$, $\Gamma^{45}$, $\Gamma^{35}$,
$\Gamma^{3}$, $\Gamma^4$ and $\Gamma^5$, which form subgroup
$\mathrm{SO(2)\times SO(4)}$ of the $\mathrm{SO(6) \sim SU(4)}$
symmetry group. More detailed analysis would show that now the GSM
is the Grassmann manifold $\mathrm{SO(6)/[SO(2)\times SO(4)]} =
G_{2,6}$ (Fig. \ref{halffilled}), which is defined as the set of 2
dimensional planes in 6 dimensional vector space.

%The half-filled Heisenberg model has a similar form as Eq.
%\ref{model}, although we will still use the notation of
%$\Gamma^{ab}$ and $\Gamma^a$, the representation on every site is
%no longer fundamental. The point $J_1 = - J_2
%> 0$ is apparently SU(4) invariant, when $|J_2| < J_1$ the phase
%obtained by mean field calculation has nonzero $\langle (-1)^i
%\Gamma^{ab}\rangle$ \cite{tu2007}, which corresponds to the state
%discussed in the previous paragraph with Grassmann manifold
%$G_{2,5}$ instead of CP(3). When $|J_2| > J_1$ the phase has
%nonzero $\langle(-1)^i \Gamma^a\rangle$, which is a staggered
%nematic state with GSM $\mathrm{SO(5)/SO(4)} = S^4$. At the SU(4)
%point, because of the enlarged symmetry, the two orders should
%have equal energy. Suppose $(-1)^i\Gamma^{12}$ is nonzero at this
%SU(4) point, now the residual symmetry of this order is generated
%by $\Gamma^{12}$, $\Gamma^{34}$, $\Gamma^{45}$, $\Gamma^{35}$,
%$\Gamma^{3}$, $\Gamma^4$ and $\Gamma^5$, which form subgroup
%$\mathrm{SO(2)\times SO(4)}$ of the $\mathrm{SO(6) \sim SU(4)}$
%symmetry group. More detailed analysis would show that now the GSM
%is the Grassmann manifold $\mathrm{SO(6)/[SO(2)\times SO(4)]} =
%G_{2,6}$ (Fig. \ref{halffilled}), which is defined as the set of 2
%dimensional planes in 6 dimensional vector space.

One can write down a GL field theory for the half-filled Sp(4)
spin system as follows: \beqn F_{hf} &=& \sum_{ab,\mu}(\nabla_\mu
\phi^{ab})^2 + (\nabla_\mu \phi^a)^2 + r_1 (\phi^{ab})^2 + r_2
(\phi^a)^2 \cr\cr &+& g\{\sum_{ab}(\phi^{ab})^2 +
\sum_a(\phi^a)^2\}^2 + \sum_a \gamma_2
(\epsilon_{abcde}\phi^{bc}\phi^{de})^2 \cr\cr &+& \cdots
\label{GLhalffilled}\eeqn $r_1 = r_2$ corresponds to the SU(4)
point, and $r_2 < r_1$ ($r_2
> r_1$) corresponds to the case with $\langle (-1)^i\Gamma^a \rangle \neq 0$
($\langle (-1)^i\Gamma^{ab} \rangle \neq 0$). Notice that the cubic term
$\gamma\epsilon_{abcde}\phi^a\phi^{bc}\phi^{de}$ is not allowed
here because $\phi^a$ and $\phi^{ab}$ both represent staggered
orders, so this cubic term would switch sign under lattice
translation. The ellipses in Eq. \ref{GLhalffilled} includes other
terms allowed by symmetry, for instance
$\sum_{ab}(\epsilon_{abcde}\phi^c\phi^{de})^2$.

In 2+1 dimensional space, another possible ground state around the
SU(4) point of the half-filled system is the algebraic spin
liquid, which has been actively studied analytically
\cite{brad1988,brad1989,Hermele2004a,Hermele2005a,wen2002a,xu2008}
and has gained numerical supports \cite{assaad2005}. However, the
fate of the SU(4) point at three dimension is unclear, so in this
work we tentatively assume it still has magnetic order which
bridges the orders on two sides of the phase diagram in Fig.
\ref{halffilled}$b$, and the transition between the two different
spin order patterns at zero temperature should be first order.

\subsection{B, Noncollinear Phases}

Now let us move on to the GL theory for Sp(4) spin system with
noncollinear spin orders. It was shown \cite{xuyang2008} that the
GSM of the ordered phase of Sp(4) system on the triangular lattice
is $S^7/Z^2$ with $\sqrt{3}\times \sqrt{3}$ order of $\Gamma^{ab}$
and collinear and uniform order of $\Gamma^a$. By tuning $J_2/J_1$
there is a transition between the ordered phase and a deconfined
$Z_2$ spin liquid which belongs to the 3d O(8) universality class.
Now let us consider the Sp(4) Heisenberg model on the stacked
triangular lattice, and study the GL theory in terms of physical
order parameters. This ordered state is characterized by the
$\sqrt{3}\times \sqrt{3}$ order of $\phi^{ab}_1 + i\phi^{ab}_2 =
z^t \Gamma^{ab} z$, and a uniform order of $\phi^a = z^\dagger
\Gamma^a z$. $z_\alpha$ is the Sp(4) bosonic spinon expanded at
the minima of the spinon band structure, which are located at the
corners of the hexagonal Brillouin zone $\vec{Q} = (\pm 4\pi/3,
0)$. The two 10 component Sp(4) adjoint vectors $\phi^{ab}_1$ and
$\phi^{ab}_2$ are ``perpendicular" to each other: $\sum_{a,b}
\phi^{ab}_1\phi^{ab}_2 = 0$. In the ordered state, The vectors
$\phi^{ab}_1$, $\phi^{ab}_2$ and $\phi^a$ satisfy the following
relations: \beqn \epsilon_{abcde}\phi^{bc}_1\phi^{de}_1 =
\epsilon_{abcde}\phi^{bc}_2\phi^{de}_2 \sim |z|^2\phi^a. \eeqn
Therefore the GL theory reads: \beqn F &=& \sum_{i = 1}^2
\sum_{a,b} (\nabla_\mu \phi^{ab}_i)^2 + r_1 (\phi^{ab}_i)^2 +
(\nabla_\mu\phi^a)^2 +  r_2 (\phi^a)^2 \cr\cr &+&
\sum_{i}\gamma\epsilon_{abcde}\phi^a\phi^{bc}_i\phi^{de}_i +
g_3[\sum_{ab,i} (\phi^{ab}_i)^2]^2 \cr\cr &+& g_4
\{(\sum_{ab}\phi_1^{ab}\phi_2^{ab})^2 - [\sum_{ab}
(\phi^{ab}_1)^2][\sum_{cd}(\phi^{cd}_2)^2] \}. \label{GL4}\eeqn
The last term in (\ref{GL4}) with $g_4
> 0$ guarantees the ``orthogonality" between $\phi^{ab}_1$ and
$\phi^{ab}_2$ in the ordered phase. Besides the apparent Sp(4)
symmetry, this free energy Eq. \ref{GL4} within the forth order
has an extra O(2) symmetry for rotation between $\phi^{ab}_1$ and
$\phi^{ab}_2$, which corresponds to the translation symmetry of
the system: \beqn T_x: \  \phi^{ab}_1 + i\phi^{ab}_2 \rightarrow
(\phi^{ab}_1 + i\phi^{ab}_2)\exp(i 2\pi/3). \eeqn For the
commensurate $\sqrt{3}\times\sqrt{3}$ order, this O(2) symmetry
will be broken by the sixth order terms of this free energy; if
the noncollinear state is incommensurate, the O(2) symmetry will
be preserved by any higher order of the GL theory.

In the GL theory Eq. \ref{GL4}, depending on $\Delta r = r_2 -
r_1$, the order of $\phi^a$ is allowed to occur before the order
of $\phi^{ab}_i$, and the transition of $\phi^a$ again belongs to
the O(5) universality class. After the order of $\phi^a$, the
quadratic part of the free energy (\ref{GL4}) can be diagonalized,
and O(3) vectors $\vec{T}_{A,1}$ and $\vec{T}_{A,2}$ would order
after $\phi^a$. The last term in (\ref{GL4}) would induce a term
$(\vec{T}_{A,1}\cdot\vec{T}_{A,2})^2$ at this transition,
therefore the field theory for the second transition is described
by the following coupled O(3) free energy \beqn F &=& \sum_{i =
1}^2(\nabla_\mu \vec{n}_i)^2 + r (\vec{n}_i)^2 + v\{(\vec{n}_1)^2
+ (\vec{n}_2)^2\}^2\cr\cr &+& u \{(\vec{n}_1\cdot \vec{n}_2)^2 -
(\vec{n}_1)^2(\vec{n}_2)^2\} +
 \cdots \label{n1n2}\eeqn with
$\vec{n}_{i} = \vec{T}_{A, i}$. Again the Goldstone mode of
$\phi^a$ only induces irrelevant perturbation. This coupled O(3)
model defined in Ref. \cite{kawamura1990} with symmetry
$\mathrm{O(2) \times O(3)}$ has attracted enormous analytical and
numerical work, recent results suggest the existence of a new
universality class of the coupled O(3) model \cite{vicari2004}.
When $n_1$ and $n_2$ are ordered, the whole SO(3) symmetry
associated with $\vec{T}_{A,i}$ is broken, and the residual
symmetry of the condensate of $\vec{n}_i$ is SO(3), which is the
SO(3) symmetry associated with $\vec{T}_{B, i}$ $i.e.$
$\mathrm{SU(2)}_B$.

Again the nature of the GSM depends on which type of SO(3) the
residual symmetry is. For half-filled spin-3/2 cold atoms on the
triangular lattice, one can engineer a state without order of
$\Gamma^{ab}$, but with $\sqrt{3}\times\sqrt{3}$ order of nematic
order parameter $\Gamma^a$: \beqn \langle \Gamma^a(\vec{r})
\rangle &\sim& n^a_1\cos(\vec{Q}\cdot\vec{r}) + i
n^a_2\sin(\vec{Q}\cdot\vec{r}), \cr \cr && \sum_{a=1}^5 n_1^an_2^a
= 0. \eeqn This spiral nematic order parameter has residual
symmetry SO(3), however this is the $\mathrm{SU(2)}_V$ subgroup
discussed previously. For instance if $\vec{n}_1 = (1, 0, 0, 0,
0)$ and $\vec{n}_2 = (0,1,0,0,0)$ then $\mathrm{SU(2)}_V$ is
generated by $\Gamma_{34}$, $\Gamma_{45}$ and $\Gamma_{35}$.
Therefore the GSM of this order can be written as quotient space
$\mathrm{SO(5)/SO(3)}$, but not equivalent to $S^7/Z_2$. The GL
theory describing this nematic $\sqrt{3}\times\sqrt{3}$ order is a
coupled O(5) sigma model, which is analogous to Eq. \ref{n1n2}.

Another state worth mentioning briefly is the superconductor state
of the Sp(4) fermions, and we will only focus on the $s-$wave
pairing here. The $s-$wave pairing of two Sp(4) particles can be
either Sp(4) singlet or quintet. And the quintet state which is
characterized by a complex O(5) vector $\vec{d} = \vec{d}_1 +
i\vec{d}_2 $ can have two types of GSM, depending on the
microscopic parameters of the system. The first type of pairing
has $\vec{d}_1$ parallel with $\vec{d}_2$, then the GSM is
$[S^4\times S^1]/Z_2$ \cite{WU2005b}. The second type of pairing
has $\vec{d}_1\cdot\vec{d}_2 = 0$, then the GSM is again
characterized by two real orthogonal O(5) vectors, and hence
GSM$=\mathrm{SO(5)/SO(3)}$, equivalent to the nematic
$\sqrt{3}\times\sqrt{3}$ state discussed in the previous
paragraph. In experimental system with spin-3/2 cold atoms, the
direct calculation with $s-$wave scattering suggests that the
former state (dubbed polar state) is likely favored
\cite{wu2006b}.

\section{3, close to quantum phase transitions}

In this section we will study the phase transitions obtained in
the previous section in the region close to a quantum phase
transition. For two dimensional square lattice, it was proposed in
Ref. \cite{xuyang2008} that by tuning $J_2/J_1$ in Eq.
\ref{model}, there is a deconfined quantum phase transition
between Neel order and a gapped plaquette order which belongs to
the 3d CP(3) universality class. If now we turn on a weak spin
interaction between square lattice layers, the deconfined quantum
phase transition is expected to expand into a stable spin liquid
phase with gapless photon excitation, while the Neel order and
plaquette order are unaltered by the weak $z$ direction
tunnelling.

\begin{figure}
\includegraphics[width=3.0in]{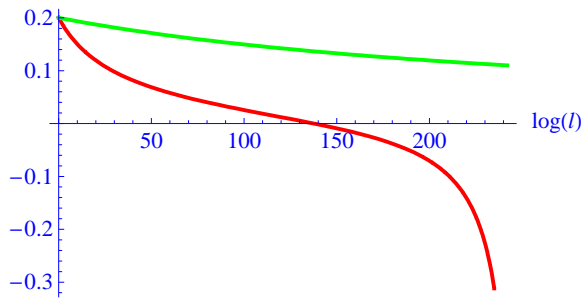}
\caption{The RG flow for $e^2$ (green line) and $g$ (red line) in
Eq. \ref{3dcp3} with trial initial value $g_0 = e^2_0 = 1/5$. }
\label{RG}
\end{figure}

The quantum phase transition between Neel and photon phase is
described by the 3+1d CP(3) model: \beqn L &=& \sum_{a = 1}^4
|(\partial_\mu - iA_\mu)z_a|^2 + r|z_a|^2 + g (|z_a|^2)^2 \cr\cr
&+& \frac{1}{16 e^2}F_{\mu\nu}^2 + \cdots \label{3dcp3}\eeqn Based
on naive power counting this 3+1d transition is a mean field
theory with marginally relevant/irrelevant perturbations. To
determine the universality class of this transition, we need to
calculate the RG equation for $g$ and $e^2$ in Eq. \ref{3dcp3} in
detail. At the transition with $r = 0$, the coupled RG equation up
to one loop for $g$ and $e^2$ reads: \beqn \frac{dg}{d\ln l} &=&
-\frac{2}{\pi^2} g^2 - \frac{3}{8\pi^2}e^4 + \frac{3}{4\pi^2}e^2g,
\cr\cr \frac{de^2}{d\ln l} &=& - \frac{1}{6\pi^2}e^4.
\label{rgeq}\eeqn The RG equation for the Higgs model with $N = 1$
was calculated in Ref. \cite{coleman1973}, the structure of the RG
equation obtained therein is quite similar to Eq. \ref{rgeq}.
Taking this RG equation, one can see that the electric charge
$e^2$ is always renormalized small. If one starts with a positive
value of $g$, $g$ will be first renormalized to smaller values
marginally, and then switch sign due to its coupling with $e^2$,
and finally becomes nonperturbative, and no fixed point is found
with arbitrary choices of initial values of $g$ and $e^2$. So
eventually this transition is probably weak first order. The
solution of RG equations Eq. \ref{rgeq} is plotted in Fig.
\ref{RG} for the trial initial value $g_0 = e_0^2 = 1/5$. One can
see that $g$ becomes nonperturbative much slower than an ordinary
marginally relevant operator, because the ordinary marginally
relevant operator will still monotonically increase under RG flow.
In our current case $g$ remains perturbative and decreases for a
very large energy scale, so for sufficiently small initial values
of $g$ and $e^2$, at physically relevant energy scale, we can
treat this transition a mean field transition of spinon
$z_\alpha$.

%For instance the classical transition critical temperature of the
%ordered state and the ordered moment at zero temperature, will
%have the following scaling close to the quantum transition: \beqn
%T_c \sim r^{1/2}, \ \ \langle \phi^{ab} \rangle \sim r. \eeqn

\begin{figure}
\includegraphics[width=2.6in]{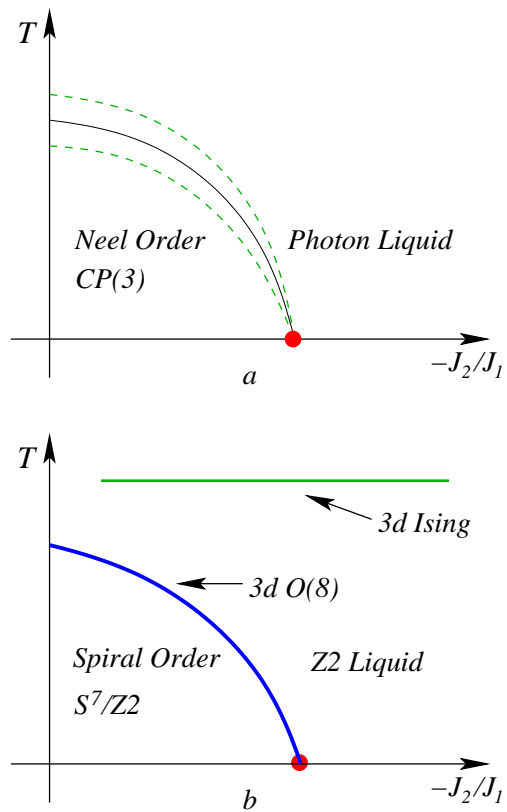}
\caption{The phase diagram close to the quantum phase transitions
in stacked square ($a$) and triangular lattices ($b$). The region
between the dashed lines in Fig. $a$ is the ``monopole dominated"
region which should be described by GL theory Eq. \ref{GL}. The
blue curve in Fig .$b$ is a 3d O(8) transition, and the green line
is a 3d Ising transition which separates a low temperature
classical $Z_2$ spin liquid from a high temperature featureless
disordered phase.} \label{sp4quantumdia}
\end{figure}

Without monopoles, the finite temperature transition will be
described by the 3d CP(3) model in Eq. \ref{cpn}. If temperature
is turned on, finite density of monopoles will be generated. Close
to the quantum transition, since the critical temperature of the
Neel order is very low, the monopoles roughly have small fugacity
$y_m \sim \exp(-E_g/T)$, and $E_g$ represents the short distance
energy gap of monopole. Therefore very close to the quantum phase
transition with small $T_c$, there is a very narrow ``monopole
dominated" region around the classical phase transition where the
universal physics significantly deviates from the CP(3) model.
Inside the monopole dominated region the GL field theory in Eq.
\ref{GL} becomes applicable, with $r = r_1 + r_2$ tuned by
temperature. Out of this monopole dominated region, the scaling
behavior of the 3d CP(3) model becomes more applicable, assuming
the noncompact CP(3) model has a second order transition. The size
of the monopole dominated region can be estimated from the
fugacity of the monopoles. If the scaling dimension of the
monopole operator at the CP(3) fixed point is $\Delta_m$, the size
of the monopole dominated range is estimated as $\Delta T/T_c \sim
y_m^{\frac{1}{(3 - \Delta_m) \nu}}$, $\nu$ is the standard
exponent of 3d CP(3) transition defined as $\xi \sim r^{-\nu}$.
The phase diagram is shown in Fig. \ref{sp4quantumdia}.

The situation is quite different for the stacked triangular
lattice. In Ref. \cite{xuyang2008} we showed that on 2d triangular
lattice, by tuning $J_2/J_1$ there is a 3d O(8) transition between
the $\sqrt{3}\times \sqrt{3}$ order and the $Z_2$ spin liquid
state, despite the fact that the microscopic system only has
Sp(4)$\sim$SO(5)$\subset$SO(8) symmetry. For a stacked triangular
lattice with weak interlayer coupling, both the
$\sqrt{3}\times\sqrt{3}$ order and the $Z_2$ spin liquid will
survive, but the quantum phase transition is described by the mean
field theory of $z_\alpha$, because the $Z_2$ spin liquid does not
introduce any critical correlation for $z_\alpha$. Notice that at
this mean field transition the magnetic order parameters
$\phi^{ab}$ will have anomalous dimension $1$, because it is a
bilinear of $z_\alpha$. For 3d space, the $Z_2$ spin liquid can
survive and extend into a finite region in the phase diagram at
finite temperature, therefore close to the quantum transition,
after the thermal fluctuation destroys the magnetic order, the
system does not enter the high temperature featureless phase
immediately, instead it enters the finite temperature $Z_2$ spin
liquid phase, and the classical transition of the spin order will
simply belong to the 3d O(8) universality class. At even higher
temperature, there is a phase transition separating the classical
$Z_2$ spin liquid state and high temperature disordered phase,
which physically corresponds to the proliferation of the ``vison
loop". This transition belongs to the 3d Ising universality class.

\section{4, Summary and outlook}

In this work we used the Ginzburg-Landau field theory to describe
and classify Sp(4) spin orders with different ground state
manifolds, and studied the nature of classical phase transitions
between these spin order and disordered phases. Our results can be
applied to Sp(4) spin models like the $J_1-J_2$ Heisenberg model
in Eq. \ref{model}. The GL theory can be generalized for large N
spin systems with GSM $\mathrm{CP(N-1)}$, for instance the cubic
term in Eq. \ref{GL} is always allowed by spin symmetry for large
N, although other discrete symmetries have to be checked
carefully.

The monopole of the gauge field $A_\mu$ will create and annihilate
the quantized flux of $A_\mu$, which equals to the soliton number
of the GSM $\mathrm{CP(N-1)}$, and the existence of soliton of
system with GSM $\mathrm{CP(N-1)}$ is due to the fact that
$\pi_2[\mathrm{CP(N-1)}] = Z$ for general N \cite{sachdev1990}. If
we start with a nonlinear sigma model for $\mathrm{CP(N-1)}$
manifold at $2+\epsilon$ dimension, though the spin wave
excitations can be quite nicely described, the expansion of
$\epsilon$ at the phase transition will not take into account of
the effect of monopoles. Therefore the $2+\epsilon$ expansion with
extrapolation $\epsilon \rightarrow 1$ is probably equivalent to
the $\mathrm{CP(N-1)}$ model in Eq. \ref{cpn}. However, if we
start with a linear sigma model at $4 - \epsilon$ dimension, the
$4 - \epsilon$ expansion will contain the information of
monopoles, and the limit $\epsilon \rightarrow 1$ is likely
converging to the true situation at three dimension. Since phase
transition is what we are most interested in, in this work we were
focusing on the linear sigma model in $4 - \epsilon$ dimension.

Another manifold which potentially can be realized by Sp(4) spin
system is the ``squashed $S^7$".  The squashed $S^7$ has been
studied for over two decades in high energy theory, as one of the
solutions of the 11 dimensional supergravity field equation is
$\mathrm{AdS}_4 \times S^7_{\mathrm{squash}}$ \cite{awada1983}.
The squashed $S^7$ is a seven dimensional manifold with the same
topology as $S^7$, but different metric and isometry group. The
ordinary $S^7$ has isometry group SO(8), and the squashed $S^7$
has isometry group $\mathrm{SO(5)\times SO(3) \subset SO(8)}$, and
the SO(5) and SO(3) commute with each other. Written as a quotient
space, the squashed $S^7$ can be expressed as \cite{bais1983}
\beqn S^{7}_{\mathrm{squash}} = [\mathrm{SO(5)} \times
\mathrm{SO(3)}_C] / [\mathrm{SO(3)}_A \times \mathrm{SO(3)}_D],
\eeqn Here $\mathrm{SO(3)}_A$ is a normal subgroup of one of the
SO(4) subgroup of the SO(5) group in the numerator, and the other
normal SO(3) subgroup of this SO(4) is denoted as
$\mathrm{SO(3)}_B$, $i.e.$ $\mathrm{SO(3)}_A \times
\mathrm{SO(3)}_B \sim \mathrm{SO(4)}$. $\mathrm{SO(3)}_D$ is the
diagonal subgroup of $\mathrm{SO(3)}_B \times \mathrm{SO(3)}_C$,
$i.e.$ $J^D_i = J^B_i + J^C_i$, $i = 1, 2, 3$. To realize the
squashed $S^7$ GSM, we should start with a system with global
symmetry $\mathrm{SO(5)\times SO(3)}$. For instance, by tuning the
two $s-$wave scattering lengths, the half-filled Hubbard model of
the Sp(4) fermions can have an extra SU(2) symmetry besides the
apparent Sp(4) flavor symmetry \cite{wu2003}. Also a Sp(4) spin
liquid theory with fermionic spinons with momentum space valley
degeneracy can have an extra SU(2) symmetry contributed by the
valley degeneracy. So both cases might be a good starting point
for realizing the squashed $S^7$ manifold. We will leave the
discussion of squashed $S^7$ to future study.

\begin{acknowledgments}

The author is supported by the Society of Fellows and Milton Funds
of Harvard University.

\end{acknowledgments}

\bibliography{sptransition1}
\end{document}